\documentstyle[12pt]{article}

\def\r3{{\cal R}_3}

\begin{document}                                                              
\begin{titlepage}
December 1993 \hfill                BNL-GM-1
\vspace{1.5cm}
\begin{center} 
{\LARGE Larger domains of disoriented chiral condensate through 
annealing}\\
\vspace{1.5cm} 
\large{
Sean Gavin\\
\vspace{0.3cm}
Physics Department, Brookhaven National Laboratory\\
P.O. Box 5000, Upton, NY 11973, USA \\
\vspace{0.3cm}
and\\
\vspace{0.3cm}
Berndt M\"uller\\
\vspace{0.3cm}
Department of Physics, Duke University\\
Durham, NC, 27708-0305, USA \\
}
\vspace{1.9cm} 
\end{center}
\abstract{Relativistic heavy ion collisions can generate metastable 
domains in which the chiral condensate is disoriented.  Nucleus-sized
domains can yield measurable fluctuations in the number of neutral and
charged pions.  We propose a scenario in which domains are `annealed'
by a dynamically evolving effective potential in the heavy ion system.
Domains of sizes exceeding 3~fm are possible in this scenario.}
\vfill
\end{titlepage}

The order of the chiral restoration phase transition in QCD for
realistic values of the up, down and strange quark masses is currently
unknown \cite{qcd,ggp2}.  If the transition is nearly second order,
Rajagopal and Wilczek \cite{raw} have speculated that the
nonequilibrium dynamics of a heavy ion collision can generate
metastable domains in which the chiral condensate is disoriented.
Such domains can provide a coherent source of pions that can exhibit
novel ``centauro-like'' \cite{cen} fluctuations of neutral and
charged pions \cite{ar,bk,bkt}.  However, the ability of the coming
ion-ion experiments at the RHIC and LHC colliders to resolve such
fluctuations critically depends on the size and energy content of
these domains.  

Numerical simulations \cite{ggp1} show that the zero-temperature
linear sigma model advocated in \cite{raw} leads to domains that are
roughly pion sized, too small to be resolved in experiments.
Nevertheless, we illustrate in this Letter that finite temperature
effects in a heavy ion collision can enhance the size of domains
relative to the zero-temperature estimates.

Rajagopal and Wilczek propose a ``quench'' scenario in which the
condensate is initially chirally symmetric as appropriate at high
temperature, but its evolution is taken to follow classical equations
of motion in the absence of a heat bath.\footnote{ This procedure is
not the standard quench in condensed matter physics, in which the
system remains in contact with a heat bath of zero temperature. We
continue to use the term ``quench'' here, however, because a $T=0$
heat bath would also effectively imply decoupling from the system
since the interactions of pions effectively disappear at zero momentum
due to the approximate chiral symmetry.}  For this quench scenario
to be realistic, the expansion rate of the heat bath needed to create
the symmetric initial state must greatly exceed the rate at which the
mean field evolves.  The opposite is more likely the case.  In
\cite{raw} and \cite{ggp1}, the $\sigma$ and $\vec\pi$ fields in the
O$(4)$ linear sigma model are taken to characterize the dynamics of
the quark antiquark condensate in two-flavor QCD.  These fields evolve
over the time scale $\sim m_\sigma^{-1}\sim 1/3$~fm \cite{ggp1}, where
$m_\sigma = 600$~MeV is the zero-temperature $\sigma$ mass.  On the
other hand, model calculations \cite{kat} indicate that chiral
symmetry is broken very late in RHIC-energy collisions, perhaps at
proper times $\tau_c \sim 5-20$~fm, so that the expansion time scale
$\sim \tau_c$ is much larger in comparison.

We present an alternative ``annealing'' scenario in which the
sigma-model condensate evolves semiclassically in the presence of a
bath of quasiparticle excitations.  Fluctuations
induced by the quasiparticles create an effective potential \cite{qft}
in which a chirally symmetric state is initially stable (modulo finite
quark-mass effects).  As the system expands and rarefies, the
effective potential develops a ``wine-bottle'' shape and slowly changes
towards its free-space depth.  The condensate then evolves in this
changing potential.

Transient domains of disoriented ${\Phi}\equiv (\sigma, {\vec
\pi})$ condensate can develop because the ${\Phi}\approx 0$
initial state is unstable in the wine bottle potential.  Consider
first a quench scenario in which the potential, 
\begin{equation}
V_0({\Phi}) = \lambda(\Phi^2 - v^2)/4 - H\sigma, 
\label{eq:pot}
\end{equation}
is fixed.  The system ``rolls down'' from the unstable local maximum
of $V_0({\Phi})$ towards the nearly stable values with $|{\Phi}| = v$,
since the symmetry breaking term $-H\sigma$ is relatively small.  This
process is analogous to spinodal decomposition in condensed matter
physics \cite{boy,bray}.  Field configurations with ${\vec\pi}\neq 0$
develop during the roll-down period.  The field will eventually settle
into stable oscillations about the unique vacuum $(f_\pi,\vec{0})$ for
$H\neq 0$, but oscillations will continue until interactions
eventually damp the motion. In the heavy-ion system a domain can
radiate pions preferentially according to its isospin content.

Domain growth in the annealing scenario with an evolving effective
potential differs from the quench as follows.  Initially, $V_{\rm
eff}$ is nearly flat for $\sigma \approx {\vec\pi}\approx 0$ as shown
in fig.~1.  Therefore, the roll-down time scale can be very large at
first, or infinite if the corresponding equilibrium transition is truly
second order (as is certainly the case for $H=0$).  Only as the
potential approaches its free space shape does the roll-down become
rapid.  In both quenching and annealing, domains can grow as long as
$\sigma$ is substantially different from $f_\pi$.  In the annealing
scenario, however, that time scale is limited by the time needed for
$V_{\rm eff}$ to approach $V_0$.  That time in turn depends on the
global dynamics of the nuclear collision.

We describe the evolution of the condensate in the context of a self
consistent Hartree-like semiclassical approach.  There, the condensate
evolves as a classical field in the presence of a fluctuating
quasiparticle bath, see {\it e.g.} \cite{bose}.  The condensate fields
${\Phi}=(\sigma,{\vec\pi})$ obey relativistic Ginzberg-Landau \cite{kss} 
equations:
\begin{equation}
(\partial^2/\partial t^2 - \nabla^2){\Phi} = \lambda\;\{v^2 -
                    3 \langle\delta\Phi_{\parallel}^2\rangle
                    - \langle\delta{\Phi}_\perp^2\rangle 
                    - |\Phi|^2\}\;{\Phi} - H n_\sigma,
\label{eq:sigma}
\end{equation}
to quadratic order in the fluctuations.  The symmetry breaking field
$H$ is fixed in the sigma direction, $n_\sigma=(1,{\vec 0})$.  Both the
condensate ${\Phi}$ and the fluctuation fields $\delta{\Phi}$
can instantaneously have any orientation in the O$(4)$ internal space.
When the $T=0$ condensate $\Phi = (f_\pi, 0)$ is realised, the
fluctuations $\delta\Phi_{\parallel}$ along $\Phi$ can
be identified with the sigma field with the familiar tree-level mass
$m_\sigma=(2\lambda f_\pi^2 + m_\pi^2)^{1/2}$.  Similarly, the three
transverse modes $\delta{\Phi}_{\perp}$ are pion-like with the
$T=0$ mass $m_\pi = (H/f_\pi)^{1/2}$.

The Hartree correction $3 \langle\delta\Phi_\parallel^2\rangle +
\langle\delta{\Phi}_{\perp}^2\rangle$ accounts for the interactions 
of sigma-like and pion-like quasiparticles with the condensate.  To
estimate this correction, we must understand the nonequilibrium
dynamics of the quasiparticles in the heavy ion system --- no trivial
task!  To simplify our exploratory treatment, we assume that the
quasiparticles are localized excitations of momentum $\vec p$ and
energy $E_{\vec p}^{\sigma,\pi}= [p^2 + (m_{\rm
eff}^{\sigma,\pi})^2]^{1/2}$, where $m_{\rm eff}^{\sigma,\pi}$ are the
effective masses of the sigma- and pion-like quasiparticles.  We can
then write the Hartree term in terms of semiclassical phase space
distributions $f_{\sigma,\pi}({\vec p},{\vec r}, t)$:
\begin{equation}
3 \langle\delta\Phi_\parallel^2\rangle 
+ \langle\delta{\Phi}_{\perp}^2\rangle 
      = 3\int d\Gamma_p {f_\sigma(p)\over E_p^\sigma}
      +  (N-1)\int d\Gamma_p {f_\pi(p)\over E_p^\pi}
\label{eq:vev}
\end{equation}
where $d\Gamma_p = d^3p/(2\pi)^3$ and $N=4$. Note that for quasiparticles 
near local thermal equilibrium, the $f_a(p)$ are Bose distributions 
so that $3 \langle\delta\Phi_\parallel^2\rangle +
\langle\delta{\Phi}_{\perp}^2\rangle\approx T^2/2$ 
for $T\gg m_{\rm eff}^\sigma$  and $m_{\rm eff}^\pi$.  
The equations of motion (\ref{eq:sigma}) with
(\ref{eq:vev}) then correspond to the more familiar effective
potential \cite{qft} shown in fig.\ 1.  The critical temperature at
which the wine bottle shape of $V_{\rm eff}$ disappears is then $T_c =
\sqrt{2}f_\pi\approx 132$~MeV. [Lattice simulations suggest that the
condensate in equilibrium QCD substantially changes at temperatures
that are somewhat higher than this O$(4)$ model estimate
\cite{qcd,qcd2}.]
  
In general, the quasiparticle distributions will satisfy
Boltzmann-Vlasov kinetic equations in which $m_{\rm eff}^{\sigma,\pi}$
are self consistent functions of the field (see, {\it e.g.}
\cite{kb}).  Here we will neglect the following: ({\it i}) the effect
of collisions among quasiparticles, and ({\it ii}) the dynamical effect of
the quasiparticle masses.  It is reasonable to neglect collisions
since, by the time $T< T_c$, the system is sufficiently dilute that
the collision frequency is small compared to the expansion rate
\cite{pions}.  On the other hand, assumption ({\it ii}) is certainly
wrong, since $m_{\rm eff}^\sigma$ must eventually change from $\sim
m_\pi$ at $T_c$ to its free-space value $\approx 600$~MeV.
\footnote
{By neglecting the time evolution of masses, we avoid the
complicated treatment in \cite{boy} that ensures that all
quasiparticle modes have effective masses satisfying $m^2 >0$.  The
present approach is adequate, since we focus on the behavior of the
condensate, and not the quasiparticles.}  We emphasize that these
approximations, while crude, allow for the most rapid ``quench''
consistent with causality, since the quasiparticles then stream away,
unimpeded by the collisions or the gain in mass.

To illustrate how the fields and quasiparticles evolve with time, we
restrict our attention for the moment to a single domain in the
interior of the collision volume.  An expanding system created in a
central Au+Au collision at RHIC will reach temperatures below
$T_c\approx 132$~MeV only after a time $\sim 5 - 20$~fm comparable to
the Au transverse radius $R_A\sim 7$~fm.  We take the quasiparticle
flow to be roughly 3-dimensional and homologous, allowing for the
fastest (causal) flow.  We assume that at this late time in the
evolution of the reaction, the condensate is invariant under radial
boosts, so that it is a function only of the proper time $\tau\equiv
(t^2 - r^2)^{1/2}$ for radial distances $r < R_A$ and late times $t >
R_A$.  The left side of (\ref{eq:sigma}) is then
\begin{equation}
\left({{\partial^2}\over{\partial t^2}} - \nabla^2
\right){\Phi} = 
\left({{\partial^2}\over {\partial \tau^2}} + {{3}\over{\tau}}
{{\partial}\over {\partial \tau}}\right){\Phi}.
\label{eq:box}
\end{equation}
Note that for the one-dimensional Bjorken expansion valid for $r,\, t
< R_A$, one would replace $\tau$ by $(t^2 - z^2)^{1/2}$ and the ``3''
multiplying the first derivative by ``1''.  However, we stress that
chiral symmetry is unbroken within that space-time region in nuclear
collisions.

Similarly, the quasiparticle distribution functions $f_{a}({\vec
p},{\vec r},t)$ for $a = \sigma,\,\pi$ are functions only of $p^\prime
\equiv (pr-E_{\vec p}t)/\tau$ and $\tau$.  Approximation ({\it i})
implies that $f(p)$ satisfies a collisionless Boltzmann equation (see
\cite{pions} and refs.\ therein) describing the free-streaming
evolution of the phase-space distribution.  We obtain $f_a({\vec
p},{\vec r},t) = f_a^c(p^\prime\tau/\tau_c)$, where the initial
distribution $f_{a}^c(p)$ at the time $\tau_c$ is assumed to have the
thermal equilibrium form $[{\rm exp}(E_p^a/T_c) - 1]^{-1}$ at the
temperature $T_c\equiv \sqrt{2}f_\pi$.  Neglecting quasiparticle
collisions, as done here, this form applies even if effective masses
are functions of the condensate ${\Phi}(\tau)$.  This solution also
applies in quasiparticle-collision dominated regime where the
expansion is adiabatic, provided that the $m_{\rm eff}^a$ can be
neglected.  The Hartree term (\ref{eq:vev}) is then a function of
$\tau$ alone.  With ({\it ii}), we estimate
\begin{equation}
(3 \langle\delta\Phi_\parallel^2\rangle 
+ \langle\delta{\Phi}_{\perp}^2\rangle) (\tau)
\approx (3 \langle\delta\Phi_\parallel^2\rangle 
        + \langle\delta{\Phi}_{\perp}^2\rangle)_c(\tau_c/\tau)^2,
\label{eq:vev2}
\end{equation}
an approximation that is adequate provided that we focus only on the
evolution of the condensate.   

We determine the average field in a single domain
${\Phi}=(\sigma,{\vec\pi})$ as a function of $\tau$ by combining
(\ref{eq:sigma}), (\ref{eq:box}) and (\ref{eq:vev2}) to
find
\begin{equation}
 \left({{\partial^2}\over {\partial \tau^2}} + {{3}\over{\tau}}
{{\partial}\over {\partial \tau}}\right) {\Phi} = \lambda v^2
\left[1 - \epsilon\, \left({{\tau_c}\over{\tau}}\right)^2\right]{\Phi} 
-\lambda|\Phi|^2{\Phi} - H n_\sigma.
\label{eq:scaling}
\end{equation}
The standard parameters $\lambda = 20$, $v = 87.4 $~MeV, and 
$H=(119\, {\rm MeV})^3$ are consistent with the values of
the pion decay constant $f_{\pi} = 92.5$~MeV and the zero temperature 
meson masses $m_{\pi} = 140$~MeV and $m_{\sigma} = 600$~MeV.  
The annealing parameter,
\begin{equation}
\epsilon\equiv v^{-2} \left(3 \langle\delta\Phi_\parallel^2\rangle_c + 
\langle\delta{\Phi}_{\perp}^2\rangle_c\right),
\label{eq:eps}
\end{equation}
determines the strength of the Hartree term.  A value $\epsilon = 1$
applies for $m_{\rm eff}^{\sigma,\pi}(T_c)\ll T_c$.  
For $m_{\rm eff}^{\sigma,\pi}\sim 140$~MeV at
$T_c\approx132$~MeV, we find $\epsilon \approx 0.5$.  The value
$\epsilon = 1$ applies if chiral restoration in an equilibrium system
occurs as a true second order phase transition.  On the other hand, a
consistent treatment at the Hartree (tadpole) level for the O$(4)$
model with $H\ne 0$ suggests $m_{\rm eff}^{\sigma,\pi} > 0$ and a
gradual, continuous chiral restoration, rather than a phase
transition.  As mentioned earlier, the case of QCD applying to the 
real world is currently ambiguous.

In Fig.\ 2a we show numerical solutions of (\ref{eq:scaling}) for
$\tau_c = 10$~fm, $\pi^0(\tau_c) = 48$~MeV, $d\pi^1/d\tau|_{\tau_c} =
95$~MeV/fm, while in Fig.\ 2b we show the case $\pi^0(\tau_c) =
5$~MeV, $d\pi^1/d\tau|_{\tau_c} =0$.\footnote{
In principle the initial conditions are determined by
microscopic fluctuations.  The initial spread in $d{\Phi}/d\tau$ and
$\Phi$ can be quite large compared to $v$.  We estimate $\langle
(d\Phi/d\tau)^2\rangle =\sum_a\int d\Gamma_p E_p^a f_c^a(p)\approx
(95\,{\rm MeV/fm})^2$ while $\langle\Phi^2\rangle\approx \sum_a\int
d\Gamma_p (E_p^a)^{-1} f_c^a(p)\approx (48 \, {\rm MeV})^2$.}  The
dark lines and light lines correspond to calculations for $\epsilon =
1$ and 0.5, respectively.  In fig 2a.\ we see that the time $\tau_f$
needed for $\sigma$ to begin to oscillate about $\sigma = f_\pi$ is
roughly $2.5~\tau_c$ for both values of $\epsilon$.  As expected,
$\tau_f$ is essentially the time needed for $V_{\rm eff}$ to fall
within roughly 20\% of the $T=0$ potential.  For $\epsilon = 0.5$, the
solution for the more quiescent initial condition shows a similar
structure, but reaches $\sigma \sim f_\pi$ sooner, at $\tau_f\approx
1.3~\tau_c$.

To study the size of domains, we now shift our focus from the average
${\Phi}(\tau)$ in a single domain to the spatial variation of the
condensate due to fluctuations in the initial fields.  Domains arise
because a spinodal instability enables small fluctuations in the
initial field configuration to grow \cite{bray}.  Neglecting the
expansion of the system for the moment, we see that small fluctuations
about ${\Phi} = 0$ in (\ref{eq:sigma}) ``run away'' as 
${\rm e}^{\Omega t}$ where $\Omega^{2}\equiv\tau_{\rm sp}^{-2} - k^2$, 
for a wave number $k <\tau_{\rm sp}^{-1}$.  
The $k=0$ mode grows the fastest, with a time scale
\begin{equation}
\tau_{\rm sp} = \left[\lambda (v^2  - 3 \langle\delta\Phi_\parallel^2\rangle
                     - \langle\delta{\Phi}_{\perp}^2\rangle\right]^{-1/2},
\label{eq:time}
\end{equation} 
which is a function of time $\tau$ itself.  In the heavy-ion system
domains grow due to the macroscopic expansion as well as the
microscopic instability of the infrared modes of the coherent chiral
condensate.    We use (\ref{eq:sigma}) to find that small fluctuations
follow
\begin{equation}
\left\{{{\partial^2}\over {\partial \tau^2}} + {{3}\over{\tau}}
{{\partial}\over {\partial \tau}} 
- {{1}\over {\tau_{\rm sp}^2}}
+ \left({{\tau_c}\over {\tau}}\right)^2 k^2\right\}{\Phi} = 0 
\label{eq:fluct}
\end{equation}
for initial conditions specified at $\tau=\tau_c$.  The factor
$(\tau_c/\tau)^2$ in front of $k^2$ in (\ref{eq:fluct}) describes the
homologous expansion of domains; an analogous term enters in
descriptions of the smoothing out of fluctuations on a local scale due
to the expansion of scales in inflationary cosmology \cite{inflation}.

To obtain a rough estimate of the growth of the domain size in the
comoving frame, we follow Boyanovsky {\it et al.} \cite{boy} 
and calculate the contribution of the growing
fluctuations to the correlation function 
\begin{equation}
\langle \pi^0({\vec r},t)\pi^0(\vec{0},t)\rangle
\; =\;  \int\! {{d^3 k}\over{2\pi^3}}\; {\rm e}^{i{\vec k}\cdot{\vec r}}
\langle\pi^0({\vec k},t)\pi^0(-{\vec k},t)\rangle,
\label{eq:corr}
\end{equation}
where $\langle\ldots\rangle$ represents an average over the fluctuating 
initial conditions.  For definiteness, we choose the $\pi^0$ direction 
in isospin, but all cartesian pion directions are equivalent by isospin 
invariance. The initial fluctuation spectrum 
$\langle \pi^0(\vec{k},\tau_c)\pi^0(-\vec{k},\tau_c)\rangle$ 
is thermal to a high degree of accuracy, yielding spatial correlations
over characteristic distances of a thermal wavelength $(\pi T)^{-1}$,
{\it i.e.} 
%
%\begin{equation}
$\langle \pi^0({\vec r},t)\pi^0(\vec{0},t)\rangle 
\sim \exp(-\pi T|{\vec r}|)$.
%\end{equation}
%
For $\{\pi T\}^{-2} < r^2 < \int\tau_{\rm sp}(\tau_c/\tau)^2 d\tau$, 
we can then evaluate (\ref{eq:fluct})
using a WKB approximation and integrate (\ref{eq:corr}) applying the
saddle-point method.  We find
\begin{equation}
\langle \pi^0({\vec r},t)\pi^0({\vec 0},t)\rangle 
\approx \langle{\pi^0(0,t)}^2\rangle \; {\rm exp} 
\left\{- r^2 \left[8\int (\tau_c/\tau)^2 \tau_{\rm sp}(\tau)   
d\tau\right]^{-1} \right\}.
\label{domain}
\end{equation}
This result suggests that domain size $R_{{}_D}$ increases roughly as
\begin{equation}
R_{{}_D}(t)^2\; \sim\; 8 \int_0^t (\tau_c/\tau)^2\tau_{\rm sp}(\tau) d\tau
\;\approx\; 8\, {\tau_{\rm sp}^0\tau_c\over\sqrt{\epsilon}}
    \left[\cos^{-1}\left(\sqrt{\epsilon}{\tau_c\over t}\right) 
- \cos^{-1}(\sqrt{\epsilon})\right]
\label{eq:size}
\end{equation}
where $\tau_{\rm sp}^0 = (\lambda v^2)^{-1/2}\approx 0.5$~fm.

Domains can grow as long as the system is unstable, {\it i.e.}  until
the $\sigma$ field begins to oscillate about the value $\sigma \sim
f_\pi$.  For a nearly critical system with $\epsilon = 1$, we estimate
this time to be $\tau_f - \tau_c\approx 1.5~\tau_c$, see figs.~2.
Such values imply that the domain size can reach $R_{{}_D}\approx
7$~fm --- very large indeed!  We obtain a more conservative estimate
$R_{{}_D}\sim 3-4$~fm by taking $\epsilon = 0.5$ and supposing that
domain growth stops after $\sigma$ reaches $f_\pi$ for the first time
at $\tau_f-\tau_c \approx (0.3-0.5)\tau_c$.  Note that the numerical
effect of the inflation factor $(\tau_c/\tau)^2$ on (\ref{eq:size}) is
insignificant for the parameter values considered.  Admittedly, numerical
simulations for the annealing scenario are needed to provide more
concrete estimates.  Such simulations are in progress \cite{ggp3}.
However, we point out that similar estimates worked surprisingly well
in describing the results of quench simulations in \cite{ggp1}.

To summarize, we have shown that the slow expansion of the matter in a
relativistic nuclear collision forms disoriented chiral condensates
through ``annealing,'' rather than ``quenching'' \cite{raw}.  Our
methods are schematic and intended only to illustrate the underlying
physics.  Nevertheless, we argue that annealing can lead to domains
significantly larger than those expected from a quench
\cite{ggp1}.  Our results therefore strengthen the suggestion
\cite{raw} that disoriented chiral condensates can be observed in
heavy ion collisions at RHIC.

%\section*{Acknowledgements}

S.G. is grateful to Andreas Gocksch and Rob Pisarski for collaboration
during the early part of this work and to Gordon Baym and Miklos
Gyulassy for discussions. B.M. thanks Chi-Yong Lin and Sergei Matinyan
for discussions during the early stage of this work.  This work was
supported in part by the Department of Energy grants DE-AC02-76CH00016
and DE-FG05-90ER40592.

\pagebreak

\section*{Figure Caption}

\noindent {\bf Figure 1}:
The Hartree effective potential as a function of
${\Phi}=(\sigma,{\vec\pi})$ for the temperatures $T_c=\sqrt{2}v$
and $T_c/3$ compared to the $T=0$ potential $V_0=\lambda(\Phi^2 -v^2)/4$.

\noindent {\bf Figure 2}:
The behavior of the field for initial conditions (a) $\pi^0(\tau_c) =
48$~MeV, $d\pi^1/d\tau(\tau_c) = 95$~MeV/fm and (b) $\pi^0(\tau_c) =
5$~MeV, $d\pi^1/d\tau(\tau_c) = 0$.  The time at which the system
first drops below $T_c$ is $\tau_c = 10$~fm.  Dark curves are calculated 
for $\epsilon=1$ and light curves for $\epsilon=0.5$.

\pagebreak

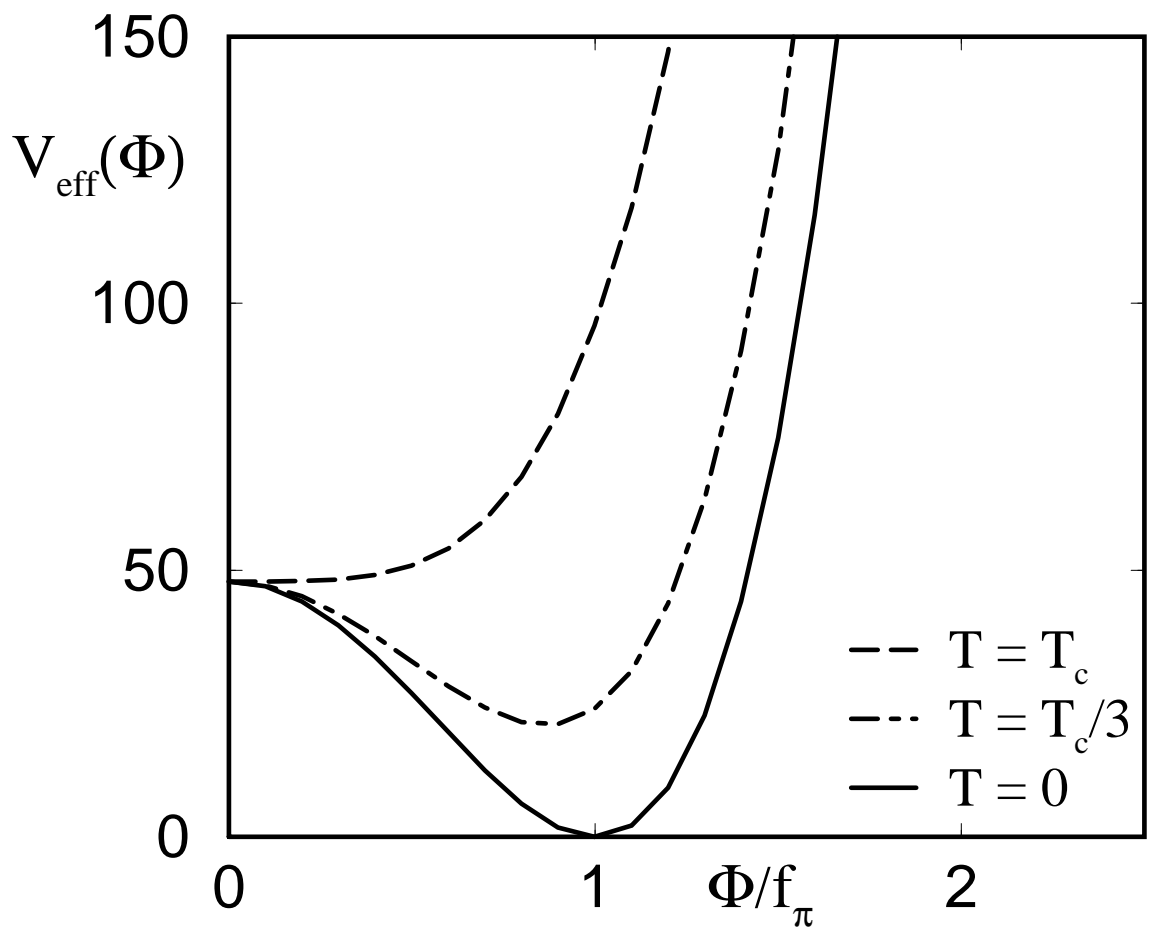
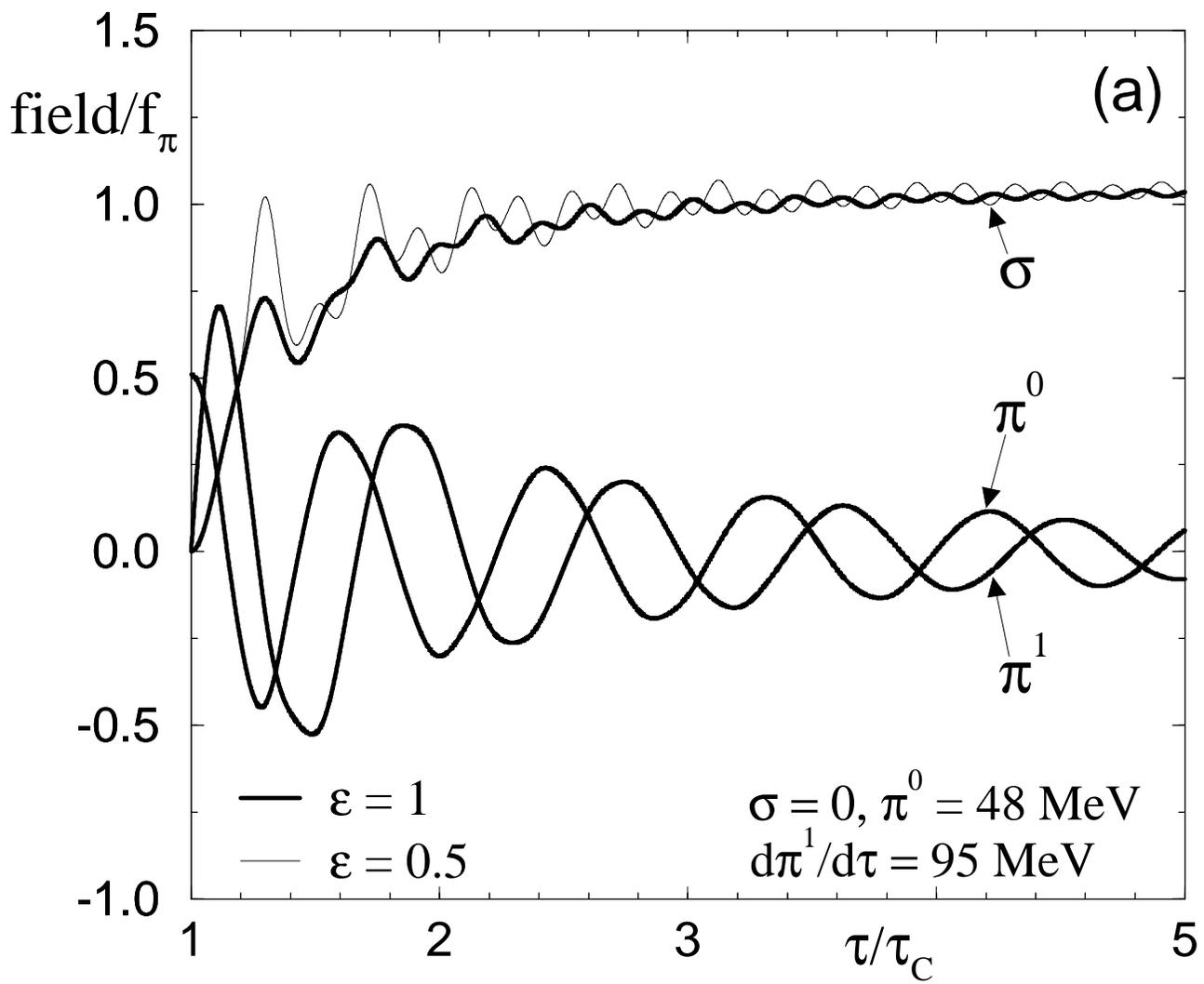
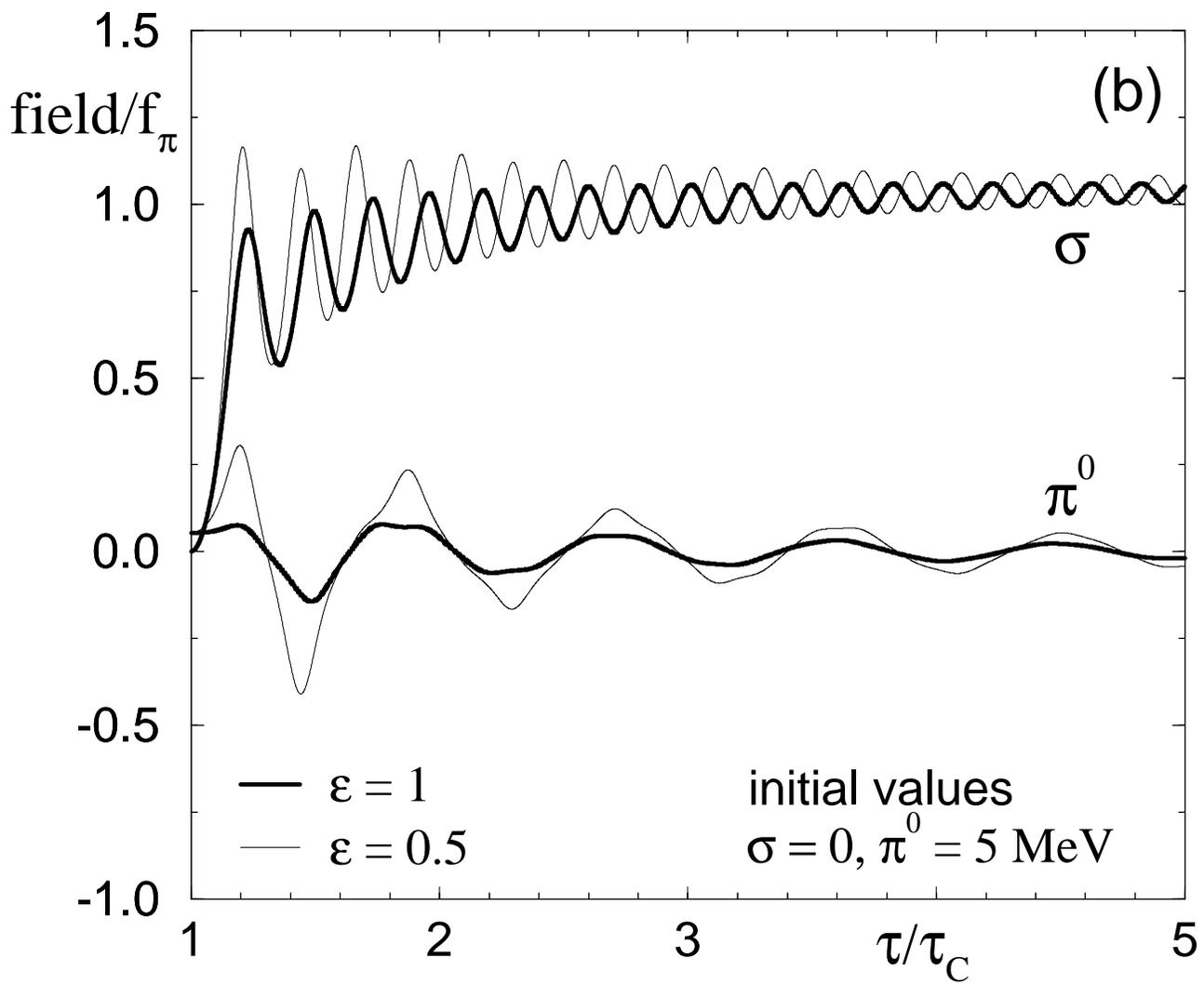
\end{document}